\documentclass[prb,amsmath]{revtex4}

\usepackage{graphicx}
\usepackage[dvips]{color}

\newcommand {\be} {\begin{equation}}
\newcommand {\ee} {\end{equation}}
\newcommand {\bea} {\begin{eqnarray}}
\newcommand {\eea} {\end{eqnarray}}
\newcommand {\bes} {\begin{displaymath}}
\newcommand {\ees} {\end{displaymath}}
\newcommand {\beas} {\begin{eqnarray*}}
\newcommand {\eeas} {\end{eqnarray*}}

\begin{document}

%%%%%%%%%%%%%%%%%%%%%%%%%%%%%%%%%%%%%%%%%%%%%%%%%%%%%%%%%%%%%%%%%%%%%
\title{Diffusion of two particles with a finite interaction potential in one dimension}

\author{Tobias Ambj\"ornsson}
\email{ambjorn@mit.edu}
\affiliation{Department of Chemistry, Massachusetts
Institute of Technology, Cambridge, MA 02139}

\author{Robert J. Silbey}
%\email{silbey@mit.edu}
\affiliation{Department of Chemistry, Massachusetts 
Institute of Technology, Cambridge, MA 02139}

\date{\today}

%%%%%%%%%%%%%%%%%%%%%%%%%%%%%%%%%%%%%%%%%%%%%%%%%%%%%%%%%%%%%%%%%%%%%

% A B S T R A C T

\begin{abstract}
We investigate the dynamics of two interacting diffusing particles in
an infinite effectively one dimensional system; the particles interact
through a step-like potential of width $b$ and height $\phi_0$ and are
allowed to pass one another. By solving the corresponding 2+1-variate
Fokker-Planck equation an exact result for the two particle
conditional probability density function (PDF) is obtained for
arbitrary initial particle positions. From the two-particle PDF we
obtain the overtake probability, i.e. the probability that the two
particles has exchanged positions at time $t$ compared to the initial
configuration. In addition, we calculate the trapping probability,
i.e. the probability that the two particles are trapped close to each
other (within the barrier width $b$) at time $t$, which is mainly of
interest for an attractive potential, $\phi_0<0$. We also investigate
the tagged particle PDF, relevant for describing the dynamics of one
particle which is fluorescently labeled. Our analytic results are in
excellent agreement with the results of stochastic simulations, which
are performed using the Gillespie algorithm.
\end{abstract}
%\pacs{}

\maketitle

%%%%%%%%%%%%%%%%%%%%%%%%%%%%%%%%%%%%%%%%%%%%%%%%%%%%%%%%%%%%%%%%%%%%%

\section{Introduction} \label{sec:Introduction}

As recent advances in manufacturing methods drives device sizes toward
the nanorange, the understanding of how interactions between diffusing
entities affect dynamics is becoming increasingly important
\cite{CD}. Situations where diffusing molecules interact strongly
are also of importance in biological systems \cite{Ellis}.

The interaction between diffusing particles can be of either attractive or
repulsive nature. For repulsive interactions, a particularly prominent
example is that of single-file diffusion, i.e. the diffusion of
identical particles which interact via a hardcore repulsion (the interaction
potential energy is plus infinity, so that the particles cannot pass
each other) in one dimension. For single-filing systems the particle
order is thus conserved over time $(t)$ resulting in interesting
dynamical behavior for a tagged particle. For instance, in contrast to
ordinary diffusion for which the mean square displacement $\langle
(x_{\cal T}-x_{{\cal T},0} )^2\rangle $ is proportional to $t$, for
single file diffusion the mean square displacement of a tagged
particle is proportional square root of time, $\langle (x_{\cal
T}-x_{{\cal T},0} )^2\rangle \propto t^{1/2}$ for long times in an
infinite system with a fix particle concentration
\cite{HA,Alexander_Pincus_78,WBL}; the probability density function
(PDF) the single-file of the tagged particle position is Gaussian
\cite{HA,LE,Beijeren_83,Hahn_95,RKH,Kollmann_03,Jara_Landim_06,Lizana_Ambjornsson}. For
attractive interactions an especially well-studied example is that of
reaction-diffusion system \cite{Evans_02}, where often the
particles are assumed to annihilate each other upon encounter,
i.e. the potential energy between particles is assumed to be minus
infinity.

Although much work has been dedicated to interacting diffusing
particles interacting via infinite (negative or positive) potentials,
to our knowledge, much fewer studies consider finite potentials.  In
 Ref. \onlinecite{Kutner_84} the problem of diffusion of $N$ particles on two
coupled linear chains was studied. Similarly, in Refs. \onlinecite{Hahn_98} and
\onlinecite{Mon_02} diffusion of spherical particles in a cylindrical
geometry, where the cylinder radius was large enough to allow passage
of particles, were studied; in Ref. \onlinecite{Hahn_98} molecular dynamics
simulations were done using a Lennard Jones interaction between
particles and in Ref. \onlinecite{Mon_02} a Monte Carlo simulation using a hard
sphere interaction was performed. Recently, the dynamics of a tagged
particle in a system consisting of particles interacting through
screened repulsive Coulomb interactions in one dimension was
investigated\cite{Nelissen}, however only through stochastic simulations.
 In this study we derive analytic results for
diffusing particles interacting through finite potentials: we solve
analytically the problem of diffusion of two particles interacting via
a finite-sized potential of finite height in one dimension, for
arbitrary initial particle positions.  Our results generalize the
single barrier results of Ref. \onlinecite{Berdichevsky} (who solved a similar
problem by Laplace transform techniques) to arbitrary initial particle
positions.

This paper has the following organization: In
Sec. \ref{sec:ProblemDef} we state the problem under consideration and
formulate the relevant equations. In Sec. \ref{sec:twopartDist} we
provide the solution of the equations for the two particle conditional
probability density function (PDF). In Sec. \ref{sec:overtake_prob} we
use the two particle PDF to obtain the overtake probability, i.e. the
probability that the two particles at time $t$ have exchanged
positions compared to the initial configuration. In
Sec. \ref{sec:trapping_prob} we calculate the trapping probability,
i.e. the probability that the two particles are trapped close to each
other (within the barrier width) at time $t$. In
Sec. \ref{sec:tagged_PDF} we obtain the PDF for one of the particles
being tagged. We compare our analytic results to stochastic
simulations using the Gillespie algorithm and find excellent
agreement. Finally, in Sec \ref{sec:Summary} we give a summary and
outlook.

%%%%%%%%%%%%%%%%%%%%%%%%%%%%%%%%%%%%%%%%%%%%%%%%%%%%%%%%%%%%%%%%%%%%% 

\section{Problem definition}\label{sec:ProblemDef}

\begin{figure}
 \includegraphics[width=14cm]{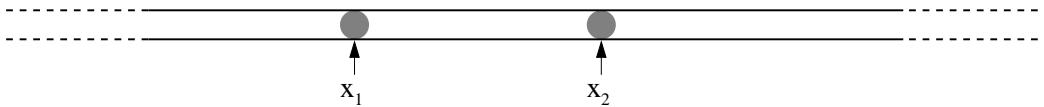}
 \caption{Cartoon of the problem considered in this study: two
   point particles are diffusing in an infinite one dimension
   system. Particle $j$ has coordinate $x_j$, and initial coordinate
   $x_{j,0}$ ($j=1,2$). The particles interact via a step-potential of
   width $b$ and height $\phi_0$, see Fig. \ref{fig:phase_space}.}
 \label{fig:cartoon}
\end{figure}

\begin{figure}
 \includegraphics[width=0.5 \textwidth]{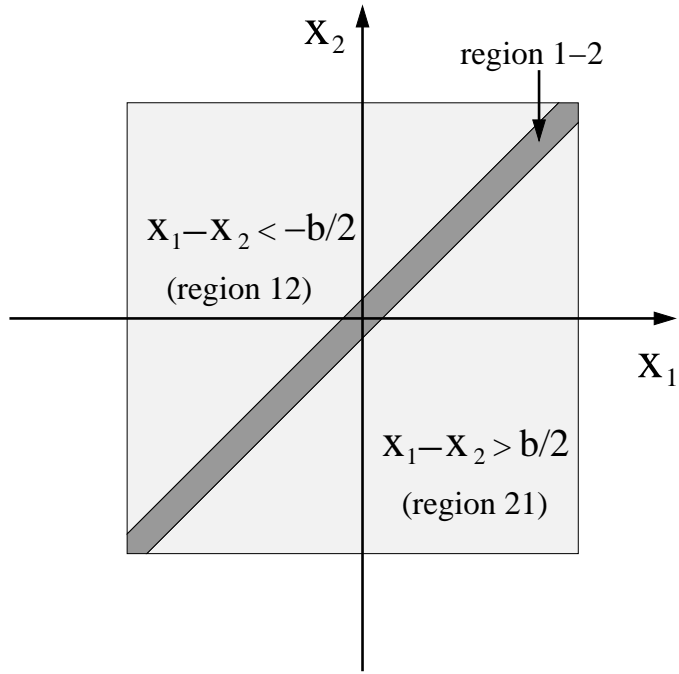}
 \caption{Phase space region for the system, the darker area,
$|x_1-x_2|<b/2$, corresponds to a region where the potential is of
height $\phi_0$, the potential width is $b$. In the lighter shaded
areas the potential is zero. The system is assumed to be infinite,
i.e. $-\infty<x_1,x_2<\infty$.}
 \label{fig:phase_space}
\end{figure}

We consider a system with two interacting point particles diffusing in an infinite one dimensional system.
The point particle problem considered here can be
transformed into a problem of {\em finite-sized} interacting particles
using a similar mapping as given in
Ref. \onlinecite{Lizana_Ambjornsson}.
A cartoon of the
system we have in mind is depicted in Fig.~\ref{fig:cartoon}.
The particles have coordinates $\vec{x}=(x_1,x_2)$ and initial
positions $\vec{x}_0=(x_{1,0},x_{2,0})$. We assume that the particles
interact through a step-like potential of width $b$ and height
$\phi_0$, i.e. the potential is
  \bea
\Phi(x_1,x_2)&=& 0 \ {\rm for} \ x_1-x_2<-b/2 \ ({\rm region} \ 12) \nonumber\\
&=&\phi_0 \ {\rm for} \ |x_1-x_2|<b/2 \ ({\rm region} \ 1-2) \nonumber\\
&=& 0 \ {\rm for} \ x_1-x_2>b/2 \ ({\rm region} \ 21).
  \eea
The phase space is depicted in Fig. \ref{fig:phase_space}, where the
darker area corresponds to the region of non-zero potential. For
$\phi_0>0$ we have a barrier, whereas for $\phi_0<0$ the potential is
of a short-range attractive nature.

The spatial distribution of the particles as a function of time is
contained in the two particle conditional PDF
${\cal P}(\vec{x},t|\vec{x}_0)$, which gives the probability of particle $1$
being in the interval $[x_1, x_1+dx_1]$ and particle $2$ in
$[x_2,x_2+dx_2]$ at time $t$ given that they initially (at time $t=0$)
were at positions $x_{10}$ and $x_{2,0}$ respectively. Since {\em
inside} each of the three regions the potential energy landscape is
flat (the force is zero) this quantity is governed by the $2+1$ (two
coordinates and time) variable Fokker-Planck (Smoluchowski) equation
\be\label{eq:DiffEq}
 \frac{ \partial {\cal P}_\gamma (\vec{x},t|\vec{x}_0)}{\partial t} = 
   D \left(
    \frac{\partial^2}{\partial x_1^2}+\frac{\partial^2}{\partial x_2^2}
  \right)
  {\cal P}_\gamma(\vec{x},t|\vec{x}_0),
\ee
where $\gamma=12,1-2$ and $21$ indicates phase-space regions, and $D$
is the diffusion constant for the particles. At the boundaries between
regions we have the following conditions \cite{Risken,Risken2}
  \bea
\label{eq:BC}
&& {\rm flux} \ J \ {\rm continuous} \ {\rm at} \ |x_1-x_2|=b/2 \nonumber\\
&& \exp(-\beta \Phi(\vec{x}) ) {\cal P}_{\gamma}(\vec{x},t|\vec{x}_0) \ {\rm continuous} \ {\rm at} \ |x_1-x_2|=b/2,
  \eea
where $\beta=1/(k_BT)$, and $k_B$ is the Boltzmann constant and $T$
the temperature. We assume that the particles start in region $12$
initially, i.e particle $1$ starts to the left of particle $2$. The
initial condition then becomes:
\bea
\label{eq:InitCond}
 {\cal P}_{12}(\vec{x},t\rightarrow 0|\vec{x}_0) &=& \delta(x_1-x_{1,0})\delta(x_2-x_{2,0}),\nonumber\\
 {\cal P}_{1-2}(\vec{x},t\rightarrow 0|\vec{x}_0) &=&  {\cal P}_{21}(\vec{x},t\rightarrow 0|\vec{x}_0)=0,
\eea
where $\delta(z)$ denotes the Dirac delta-function.

%%%%%%%%%%%%%%%%%%%%%%%%%%%%%%%%%%%%%%%%%%%%%%%%%%%%%%%%%%%%%%%%%%%%%

\section{Two-particle probability density function}\label{sec:twopartDist}

In order to solve the equations specified in the previous section we
make a variable transformation to the center-of-mass position $X$ and
relative coordinate $s$ according to: 
  \bea
\label{eq:Xs}
X&=&\frac{1}{2}\left(x_1 + x_2\right),\nonumber\\
s&=&x_1-x_2.
  \eea
Eqs. (\ref{eq:DiffEq}) and (\ref{eq:InitCond}) then becomes (leaving
the argument corresponding to the initial positions implicit)
  \bea
\label{eq:P^X^s_eq}
\frac{\partial {\cal P}_\gamma(X,s,t)}{\partial t}&=&\left(D^X \frac{\partial^2}{\partial X^2}+D^s \frac{\partial^2}{\partial s^2}\right) {\cal P}_\gamma (X,s,t),\nonumber\\
{\cal P}_{12}(X,s,t\rightarrow 0)&=& \delta(X-X_0)\delta(s-s_0),\nonumber\\
{\cal P}_{1-2}(X,s,t\rightarrow 0)&=& {\cal P}_{21}(X,s,t\rightarrow 0)= 0,
  \eea
where $X_0=(x_{1,0} + x_{2,0})/2$ and $s_0=x_{1,0}-x_{2,0}$ ($s_0<-b/2$)
and the effective diffusion constants
  \bea
\label{eq:D_eff}
D^X&=&\frac{D}{2},\nonumber\\
D^s&=&2D.
  \eea
The equations above express the fact that the relative coordinate $s$
diffuses with a diffusion constant $2D$, whereas the center-of-mass
coordinate $X$ diffuses with a diffusion constant $D/2$. The boundary
conditions, Eqs. (\ref{eq:BC}), give rise to the four equations:
  \bea
\label{eq:BC1}
\frac{\partial {\cal P}_{12}(X,s,t)}{\partial s}|_{s=-b/2}&=&\frac{\partial {\cal P}_{1-2}(X,s,t)}{\partial s}|_{s=-b/2},\nonumber\\
\frac{\partial {\cal P}_{1-2}(X,s,t)}{\partial s}|_{s=b/2}&=&\frac{\partial {\cal P}_{21}(X,s,t)}{\partial s}|_{s=b/2},
  \eea
which expresses the continuity of flux and 
  \bea
\label{eq:BC2}
&&{\cal P}_{12}(X,s=-b/2,t)=\exp(\frac{\phi_0}{k_BT}) {\cal P}_{1-2}(X,s=-b/2,t),\nonumber\\
&&\exp(\frac{\phi_0}{k_BT}) {\cal P}_{1-2}(X,s=b/2,t)={\cal P}_{21}(X,s=b/2,t),
  \eea
which corresponds to a detailed balance-like condition
\cite{van_Kampen} at the boundaries.
The boundary conditions above are different from the corresponding
problem in quantum mechanics \cite{CohenTannoudji} where the
wavefunction and the derivative of the wavefunction are continuous -
the origin of this difference is due to the fact that in the quantum
mechanical problem the potential enters in the equation of motion (the
Schr\"odinger equation), whereas in the Fokker-Planck equation it is the
force which enters.

Eqs. (\ref{eq:P^X^s_eq}), (\ref{eq:BC1}) and (\ref{eq:BC2}) allow a
product solution of the form
  \be
{\cal P}_\gamma (X,s,t)={\cal P}^X(X,t){\cal P}_\gamma^s(s,t),\label{eq:P^X^s}
  \ee
where (the boundary conditions involve only the relative coordinate $s$) 
  \be
{\cal P}^X(X,t)=\frac{1}{(4\pi D^X t)^{1/2}}\exp(-\frac{(X-X_0)^2}{4D^Xt}).\label{eq:P^X}
  \ee
For the solution for ${\cal P}_\gamma^s(s,t)$ we take a general function that satisfies the diffusion equation in each of the regions and that satisfies the boundary conditions:
  \bea
\label{eq:ansatz}
{\cal P}^s(s,t)=\int_{-\infty}^\infty \frac{dQ}{2\pi} e^{-E(Q)t}e^{-iQs_0}&\times & [C_{12}^+ e^{iQs}+C_{12}^-e^{-iQs}] \ {\rm for}\ s<-b/2\nonumber\\
&\times & [C_{1-2}^+ e^{iQs}+C_{1-2}^-e^{-iQs}] \ {\rm for}\ |s|<b/2\nonumber\\
&\times & C_{21}^+e^{iQs} \ {\rm for}\ s>b/2.
  \eea
The prefactors $C^{\pm}_\gamma$ are dependent on Q in
general. Inserting Eq. (\ref{eq:ansatz}) into the equation
of motion [see Eq. (\ref{eq:P^X^s_eq})] gives
the dispersion relation:
  \be\label{eq:disp_rel}
E(Q)=D^s Q^2.
  \ee
We proceed by setting $C_{12}^+=1$ (we will show that with this
choice the initial condition is satisfied). Inserting Eq. (\ref{eq:ansatz})
into the boundary conditions Eqs. (\ref{eq:BC1}) and (\ref{eq:BC2})
produces 4 equations for the four unknowns $C_{12}^-$, $C_{1-2}^+$ and
$C_{1-2}^-$ and $C_{21}^+$. Solving this set of equations gives:
  \bea \label{eq:coeff}
C_{12}^-(Q)&=&F \frac{1}{1-F^2 e^{2iQb}} [e^{-iQb}-e^{iQb}],\nonumber\\
C_{1-2}^+(Q)&=&(1-F)\frac{1}{1-F^2 e^{2iQb}},\nonumber\\
C_{1-2}^-(Q)&=&-F(1-F)\frac{1}{1-F^2 e^{2iQb}} e^{iQb},\nonumber\\
C_{21}^+(Q)&=&(1+F)(1-F) \frac{1}{1-F^2 e^{2iQb}},
  \eea
where we introduced an effective ``reflection'' coefficient
  \bea\label{eq:F}
F=\frac{e^{\beta \phi_0}-1}{e^{\beta \phi_0}+1}=\tanh[\frac{\beta \phi_0}{2}].
  \eea
Notice that for the case of a barrier-like potential, $\phi_0\ge 0$,
we have $0\le F<1$, where $F=0$ corresponds to the absence of the
barrier and $F\rightarrow 1$ corresponds to an infinite barrier. For
the case of an attractive potential, $\phi_0<0$, we have $-1< F <0$,
where $F\rightarrow -1$ corresponds to an infinite potential
well. Combining Eq. (\ref{eq:ansatz}), (\ref{eq:disp_rel}) and
(\ref{eq:coeff}) we have:
  \bea
\label{eq:P_s}
{\cal P}_{12}^s(s,t)&=&\Psi(s-s_0,t)+F\{ \Upsilon(-[s+s_0+b],t)-\Upsilon(-[s+s_0-b],t) \},\nonumber\\
{\cal P}_{1-2}^s(s,t)&=&(1-F)\Upsilon(s-s_0,t)-F(1-F) \Upsilon(-[s+s_0-b],t), \nonumber\\
{\cal P}_{21}^s(s,t)&=&(1-F^2)\Upsilon(s-s_0,t),
  \eea
where
  \be
\label{eq:Psi}
\Psi(\alpha,t)=\int_{-\infty}^\infty \frac{dQ}{2\pi}e^{-D^s Q^2 t}e^{i\alpha Q}= \frac{1}{(4\pi D^s t)^{1/2}} \exp \large( -\frac{\alpha^2}{4D^st}\large) 
  \ee
and
  \be
  \Upsilon(\alpha,t)=\int_{-\infty}^\infty \frac{dQ}{2\pi} \frac{1}{1-F^2 e^{2iQb}} e^{-D^s Q^2 t}e^{i\alpha Q}.
  \ee
The above expression for $\Upsilon(\alpha,t)$ can be explicitly
evaluated: using $1/(1-z)=\sum_{n=0}^\infty z^n$ (valid for $|z|<1$)
we have
  \bea 
\label{eq:Upsilon}
 \Upsilon(\alpha,t)&=&\int_{-\infty}^\infty \frac{dQ}{2\pi} \sum_{n=0}^\infty F^{2n} e^{-D^sQ^2 t}e^{i\alpha Q(\alpha+2bn)}\nonumber\\
&=& \sum_{n=0}^\infty F^{2n}\frac{1}{(4\pi D^s t)^{1/2}} \exp \large( -\frac{(\alpha+2bn)^2}{4D^st}\large) .
  \eea
i.e. $\Upsilon(\alpha,t)$ is a sum of shifted Gaussians weighted by
powers of the reflection coefficient $F$ [see Eq. (\ref{eq:F})].  The
full solution to the problem is specified by Eqs. (\ref{eq:P^X^s})
(\ref{eq:P^X}), (\ref{eq:P_s}), (\ref{eq:Psi}) and (\ref{eq:Upsilon}),
where $X$ and $s$ are related to $x_1$ and $x_2$ using
Eq. (\ref{eq:Xs}). Finally, returning to our original coordinates we
find that the two-particle conditional PDF becomes:
  \bea
\label{eq:P_explicit}
{\cal P}_{12}(\vec{x},t|\vec{x}_0)&=&Q(x_1-x_{1,0})Q(x_2-x_{2,0}) \nonumber\\
& &+ F \sum_{n=0}^\infty F^{2n} \{ Q(x_1-x_{2,0}+\frac{b}{2}-bn)Q(x_2-x_{1,0}-\frac{b}{2}+bn) \nonumber\\
& &\hspace{1.5cm}-Q(x_1-x_{2,0}-\frac{b}{2}-bn)Q(x_2-x_{1,0}+\frac{b}{2}+bn) \} \nonumber\\
{\cal P}_{1-2}(\vec{x},t|\vec{x}_0)&=&(1-F)\sum_{n=0}^\infty F^{2n} \{ Q(x_1-x_{1,0}+bn)Q(x_2-x_{2,0}-bn)\nonumber\\
& & \hspace{1.5cm}-F Q(x_1-x_{2,0}-\frac{b}{2}-bn)Q(x_2-x_{1,0}+\frac{b}{2}+bn) \} \nonumber\\
{\cal P}_{21}(\vec{x},t|\vec{x}_0)&=&(1-F^2) \sum_{n=0}^\infty F^{2n} Q(x_1-x_{1,0}+bn) Q(x_2-x_{2,0}-bn), 
  \eea
where
  \be
Q(\alpha)=\frac{1}{(4\pi Dt)^{1/2}}\exp (\frac{-\alpha^2}{4Dt}).
  \ee
It is a straightforward matter to show, using $\delta(x-a)={\rm lim}
_{t\rightarrow 0} \exp[-(x-a)^2/(4Dt)]/\sqrt{4\pi Dt}$, that the
result above satisfies the initial condition specified in Eq.
(\ref{eq:P^X^s_eq}) [and that, therefore, indeed taking $C^+_{12}=1$
was the correct choice]. In the absence of a barrier, $F\rightarrow
0$, we find ${\cal P}(x_1,x_2,t)=Q(x_1-x_{1,0})Q(x_2-x_{2,0})$ as it
should. For the case of an infinite barrier, $F\rightarrow 1$, we
have: ${\cal
P}_{12}(x_1,x_2,t)=Q(x_1-x_{1,0})Q(x_2-x_{2,0})+Q(x_1-x_{2,0}+b/2)Q(x_2-x_{1,0}-b/2)$
and ${\cal P}_{1-2}(x_1,x_2,t)={\cal P}_{21}(x_1,x_2,t)=0$ in
agreement with the result for hardcore interacting particles of linear
size $b/2$. \cite{Lizana_Ambjornsson} For the case of an infinite
potential well $F\rightarrow -1$, the solution for region 12 takes the
form: ${\cal
P}_{12}(x_1,x_2,t)=Q(x_1-x_{1,0})Q(x_2-x_{2,0})-Q(x_1-x_{2,0}+b/2)Q(x_2-x_{1,0}-b/2)$;
 this result agrees with previous results for
two vicious walkers (the random walkers kill each other upon
encounter). \cite{Fisher_84,Novotny}

\section{Overtake probability}\label{sec:overtake_prob}

In this section we calculate the overtake probability ${\cal }W(t)$, i.e. the
probability that particle 1 is to the right of particle 2 at time
$t$. We have:
  \be
{\cal }W(t)=\int dx_1 \int dx_2|_{{\rm region} \ 21} {\cal P}(x_1,x_2,t).
  \ee
Changing coordinates to $X$ and $s$, see Eq. (\ref{eq:Xs}), we get
  \bea
\label{eq:W_t}
{\cal }W(t)&=&\int_{-\infty}^{\infty} dX {\cal P}^X(X,t) \int_{b/2}^\infty ds {\cal P}_{21}^s(s,t) \nonumber\\
&=&\frac{1}{2}(1-F^2)\sum_{n=0}^\infty F^{2n} {\rm erfc}[\frac{b(2n+1/2)-s_0}{\sqrt{4D^s t}}],
  \eea
where ${\rm erfc} (z)=1-{\rm erf}(z)$ is the complementary error
function, ${\rm erf}(z)=(2/\sqrt{\pi}) \int_0^z dy \exp(-y^2)$ is the
error function \cite{ABST}, and $s_0=x_{1,0}-x_{2,0}$ as before. We have
above used the fact that $\int_{-\infty}^{\infty} dX {\cal
P}^X(X,t)=1$ [see Eqs. (\ref{eq:P^X})] together with
Eqs. (\ref{eq:P^X^s}), (\ref{eq:P_s}), (\ref{eq:Psi}) and
(\ref{eq:Upsilon}). The result given in Eq. (\ref{eq:W_t}) agrees
with the result derived in Ref.  \onlinecite{Berdichevsky} (using a
Laplace-space formalism), where the problem of passage of {\em one}
particle (since the center-of-mass coordinate is integrated out above,
${\cal }W(t)$ is effectively a one-dimensional quantity) across one
and two barriers in finite and infinite one dimensional systems were
considered - however in Ref.  \onlinecite{Berdichevsky} the initial position
were taken to be right at the left border (i.e. $s_0=-b/2$), the
result given in Eq. (\ref{eq:W_t}) thus generalizes the one
barrier (infinite system) result in Ref.  \onlinecite{Berdichevsky} to
general initial condition. We point out that ( $\phi_0 \leftrightarrow
- \phi_0$ is equivalent to $F\leftrightarrow -F$) since ${\cal }W(t)$
contain only even powers of $F$ the overtake probability is invariant
under the reversal of sign of the potential $\phi_0 \leftrightarrow
-\phi_0$; it thus takes equally long times to pass a barrier of height
$\phi_0$ as it takes to pass a trap of height $-\phi_0$. Using the
fact that ${\rm erfc}(0)=1$ we find that for long times $t\rightarrow
\infty$ we have $W(t)\rightarrow 1/2$ as it should.

\begin{figure}
 \includegraphics[width=0.6 \textwidth]{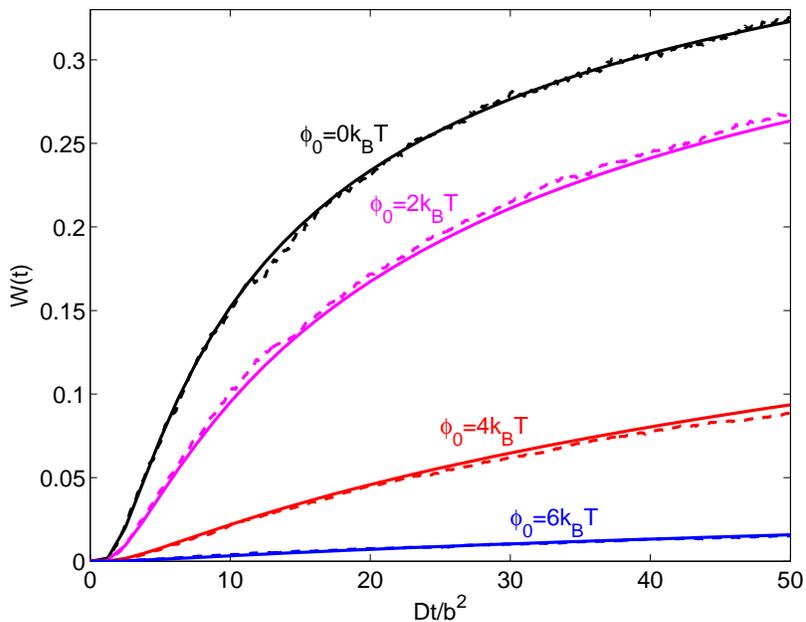}
 \caption{Overtake probability ${\cal }W(t)$ as a function of time $t$
for different barrier heights $\phi_0$. The solid curves are the
analytic result, Eq. (\ref{eq:W_t}), and the dashed curves are the
results of stochastic simulations. The following parameters were used
$x_{1,0}=-3b$, $x_{2,0}=3b$, where $b$ is the barrier width. The
simulation was performed using the Gillespie algorithm on a lattice
with $M=5000$ lattice sites, averaged over $n=10000$ ensembles,
and with a lattice spacing $a=0.025b$.}
 \label{fig:overtake_prob}
\end{figure}
In Fig. \ref{fig:overtake_prob} we illustrate the overtake
probability as given in Eq. (\ref{eq:W_t}). We notice that an
increased potential height leads to decreased probability for 
overtaking. Also, due to the time it takes for the particles to
approach each other through diffusion, there is an initial time
$\tau_c$ before the probability of overtaking becomes appreciable (even
for zero potential); a simple estimate gives $\tau_c\sim
(x_{2,0}-x_{1,0}-b)^2/(2D^s)$. In Fig. \ref{fig:overtake_prob} we also
compare the analytic results to that of a stochastic simulation using
the Gillespie algorithm (see Appendix \ref{sec:Gillespie}), and find
excellent agreement.

\section{Trapping probability}\label{sec:trapping_prob}

In this section we calculate the trapping probability ${\cal T}(t)$, i.e. the
probability that the two particles are at a distance closer than $b$
to each other at time $t$. We have
  \be
{\cal T}(t)=\int dx_1 \int dx_2|_{{\rm region} \ 1-2} {\cal P}(x_1,x_2,t).
  \ee
Again, changing coordinates to $X$ and $s$, see Eq. (\ref{eq:Xs}), we have
  \bea
\label{eq:T_t}
{\cal T}(t)&=&\int_{-\infty}^{\infty} dX {\cal P}^X(X,t) \int_{-b/2}^{b/2} ds {\cal P}_{1-2}^s(s,t), \nonumber\\
&=& \frac{(1-F)}{2} \sum_{n=0}^\infty F^{2n} \{ (1+F){\rm erf}[\frac{b(2n+1/2)-s_0}{\sqrt{4D^s t}}]\nonumber\\
&&-{\rm erf}[\frac{b(2n-1/2)-s_0}{\sqrt{4D^s t}}]-F{\rm erf}[\frac{b(2n+3/2)-s_0}{\sqrt{4D^s t}}] \}.
  \eea
For long times $t\rightarrow \infty$ we have that the trapping
probability approaches zero ${\cal T} (t)\rightarrow 0$; this is due to
ergodicity [each point in our system is assigned a probability
proportional to $\exp(-\beta \phi_0$)] combined with the fact that our
potential well is of finite width ($=b$) and connected to an infinite
system. 
\begin{figure}
 \includegraphics[width=0.6 \textwidth]{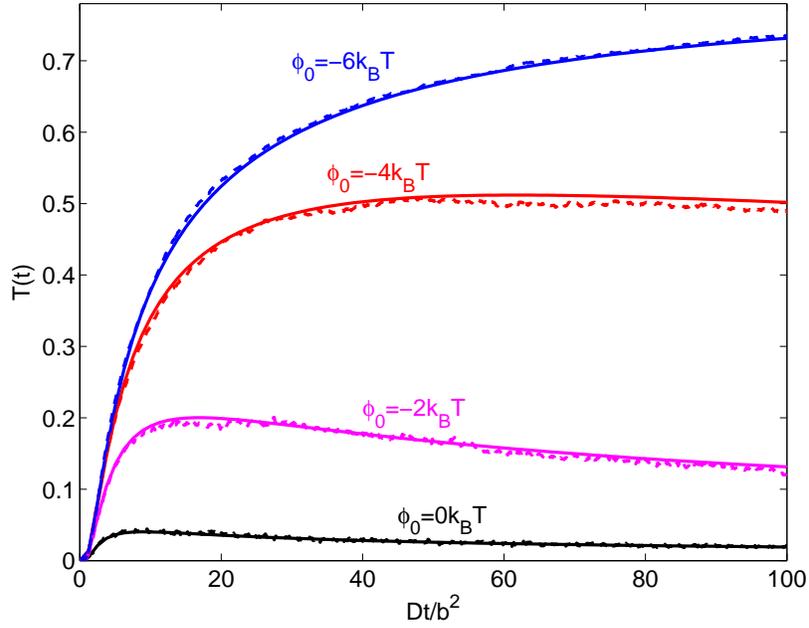}
 \caption{Trapping probability ${\cal T}(t)$ as a function of time $t$
for different potential heights $\phi_0$. The solid curves are the
analytic result, Eq. (\ref{eq:T_t}), and the dashed curves are
the results of a stochastic simulation. Except for the differently
chosen values for $\phi_0$ we used the same parameters as in Fig.
\ref{fig:overtake_prob}. }
 \label{fig:trapping_prob}
\end{figure}
In Fig. \ref{fig:trapping_prob} we illustrate the trapping
probability given by Eq. (\ref{eq:T_t}); we find excellent agreement
with stochastic simulations. We notice that a more negative
potential depth leads to increased probability for trapping as it
should. Similarly to the overtake probability, due to the time it
takes for the particles to approach each other through diffusion,
there is an initial time $\tau_c$ before any significant number of
trapping events has occurred.

A possible experiment testing the predictions in this section would
involve, for instance, two fluorescent molecules; the two molecules
interact through a potential and assuming that when the molecules are
within some distance $b$ from each other the total fluorescence get
quenched or enhanced one could directly detect a binding event between
the two particles as an increase or decrease in total fluorescence
(fluorescence measurement are here assumed ensemble averaged over
thermal noise, with fixed initial particle positions). In such an
experiment it may be more convenient to, rather than obtain ${\cal
T}(t)$ from experimental data, use time-integrated fluorescences,
i.e., measure the time-averaged trapping probability as given by
${\cal T}_{\rm av}(t)=(1/t)\int_0^t {\cal T}(t')dt'$. With the help of
Eq. (\ref{eq:T_t}) we straightforwardly obtain:
  \bea\label{eq:T_av}
  {\cal T}_{\rm av}(t)&=& \frac{(1-F)}{2} \sum_{n=0}^\infty F^{2n} \{ (1+F)G(\frac{b(2n+1/2)-s_0}{\sqrt{4D^s t}})\nonumber\\
&&-G(\frac{b(2n-1/2)-s_0}{\sqrt{4D^s t}})-FG(\frac{b(2n+3/2)-s_0}{\sqrt{4D^s t}}) \}.
  \eea
where we defined a function $G(z)=2z\exp(-z^2)/\sqrt{\pi}-2z^2{\rm
erfc}(z)+{\rm erf}(z)$.
\begin{figure}
 \includegraphics[width=0.6 \textwidth]{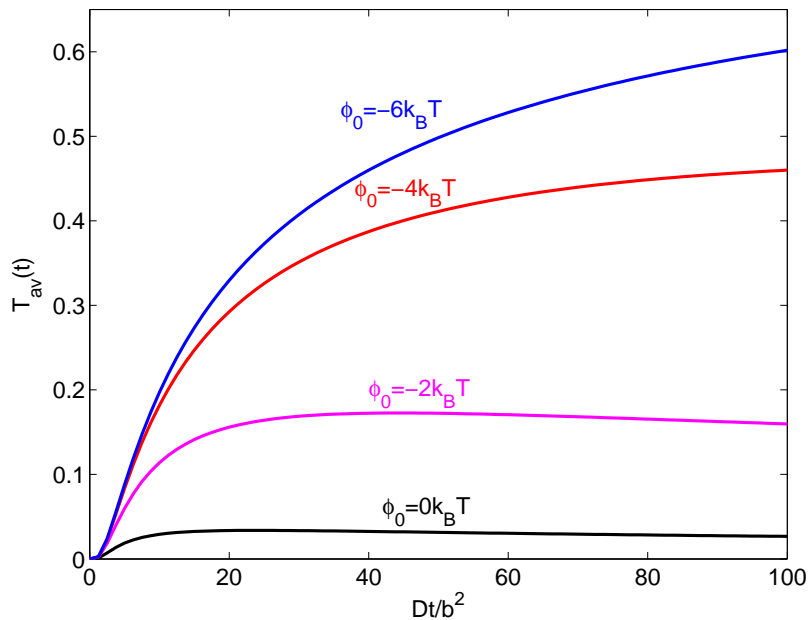}
 \caption{Time-averaged trapping probability ${\cal T}_{\rm av}(t)$ as a
function of time $t$ for different potential heights $\phi_0$,
Eq. (\ref{eq:T_av}).  Except for the differently chosen values for
$\phi_0$ we used the same parameters as in Fig.
\ref{fig:overtake_prob}. }
 \label{fig:av_trapping_prob}
\end{figure}
In Fig. \ref{fig:av_trapping_prob} we illustrate the time-averaged trapping
probability given by Eq. (\ref{eq:T_av}) for different potential
heights as a function of $t$. Experimental measurements of ${\cal
T}(t)$ or ${\cal T}_{\rm av}(t)$ would provide detailed information
about the nature of the interaction potential between the two
particles, i.e. of $\phi_0$ and $b$, by comparison to
our analytic expressions.

Finally, as a simple check of the results in Secs. \ref{sec:overtake_prob} and
\ref{sec:trapping_prob}, we calculate the first passage time
density for two vicious walkers ($F=-1$). The “survival” probability
$S(t) $ (i.e. the probability that particle 1 and 2 has not met at
time $t$) is given by $S(t) = 1- W(t)-T(t)$, which for $F = -1$
becomes $S(t) = {\rm erf}[-(s_0+b/2)/\sqrt{4D^st}] $, using
Eqs. (\ref{eq:W_t}) and (\ref{eq:T_t}). The first passage time density
is then $-dS(t)/dt=[|s_0+b/2|/\sqrt{4 \pi D^s}] t^{-3/2} \exp[
-(s_0+b/2)^2/(4D^st)]$, which agrees with the standard result
\cite{van_Kampen} for the first passage time problem for a diffusing
particle (diffusion constant $D^s$) starting a distance $s_0+b/2$ from
a perfectly absorbing wall, as it should.

\section{Tagged particle probability density}\label{sec:tagged_PDF}

By integrating out $x_2$ we can obtain the tagged particle PDF
 $\rho_1(x_1,t|\vec{x}_0)$ for particle 1.  We have:
  \bea\label{eq:tagged_pdf}
\rho_1(x_1,t|\vec{x}_0)&=&\int_{-\infty}^\infty dx_2 {\cal P}(x_1,x_2,t)\nonumber\\
&=&\int_{x_1+b/2}^\infty dx_2 {\cal P}_{12}(x_1,x_2,t) + \int_{x_1-b/2}^{x_1+b/2}dx_2 {\cal P}_{1-2}(x_1,x_2,t) \nonumber\\
&&+\int_{-\infty}^{x_1-b/2}dx_2 {\cal P}_{21}(x_1,x_2,t),
  \eea
and, using Eq. (\ref{eq:P_explicit}), we find the explicit expression
  \bea\label{eq:P1}
\hspace{-2cm}\rho_1 (x_1,t|\vec{x}_0)&=& Q(x_1-x_{1,0})R(x_1-x_{2,0}+\frac{b}{2}) \nonumber\\
& & + F \sum_{n=0}^\infty F^{2n} [ Q(x_1-x_{2,0}+\frac{b}{2}-bn) R(x_1-x_{1,0}+bn) \nonumber\\
& &\hspace{1.5cm} -F Q(x_1-x_{2,0}-\frac{b}{2}-bn) R(x_1-x_{1,0}+b+bn) ]\nonumber \\
& & + (1-F)\sum_{n=0}^\infty F^{2n} \{ -Q(x_1-x_{1,0}+bn) R(x_1-x_{2,0}+\frac{b}{2}-bn)\nonumber\\
& & \hspace{1.5cm} -F Q(x_1-x_{2,0}-\frac{b}{2}-bn)R (x_1-x_{1,0}+bn)  \nonumber\\
&&+Q(x_1-x_{1,0}+bn)[(1+F)-F R(x_1-x_{2,0}-\frac{b}{2}-bn)] \}, 
  \eea
where we introduced the function: $R(\alpha)={\rm erfc}
[\alpha/(\sqrt{4Dt})]/2$. The corresponding result for the PDF for
particle 2 is obtained by the replacements $x_1\leftrightarrow -x_2$
and $x_{1,0}\leftrightarrow -x_{2,0}$ in Eq. (\ref{eq:P1}).
In the absence of a barrier, $F\rightarrow 0$, we find
$\rho_1(x_1,t|\vec{x}_0)=Q(x_1-x_{1,0})$, i.e. the tagged particle PDF
is that of an independent diffusing particle, as it should. For the
case of an infinite barrier, $F\rightarrow 1$, we have:
$\rho_1(x_1,t|\vec{x}_0)=Q(x_1-x_{1,0})R(x_1-x_{2,0}+b/2)+Q(x_1-x_{2,0}+b/2)R(x_1-x_{1,0})$
in agreement with the result for hardcore interacting particles of
linear size $b/2$. \cite{Ambjornsson_Lizana_Silbey,Lizana_Ambjornsson}
The tagged particle PDF calculated here can experimentally be
investigated, for instance, by fluorescently labeling of one of the
particles.

\section{Summary and outlook}\label{sec:Summary}

We have in this study solved exactly the problem of diffusion of two
particles interacting via a step-like potential of height $\phi_0$ and
finite width $b$ in an infinite one-dimensional system. In particular,
from our exact analytic expression for the two particle probability
density function, we obtained the overtake probability (the
probability that the two particles has exchanged positions at time
$t$), Eq. (\ref{eq:W_t}), the trapping probability (i.e. the
probability that the particles are at distances closer than $b/2$ at
time $t$), Eq. (\ref{eq:T_t}), and the tagged particle
probability density function, Eq. (\ref{eq:P1}).

For the case of a positive potential, $\phi_0>0$ our results are of
interest for the diffusive dynamics of repulsive particles. For
instance, our results (and future extension to many particles) will be
of interest for understanding the recent simulation results dealing
with the diffusive dynamics of a tagged particles in a system of
charged particles (of the same charge) interacting through screened
Coulomb interaction \cite{Nelissen}. 

For the case of an attractive potential $\phi_0<0$, our results should
be of interest for understanding reaction-diffusion systems, where the
reaction potential between the interacting species is of finite height
and width.

It remains a future challenge to extend the results of this study to
many particles; in particular, it will be interesting to see how the
$t^{1/2}$ (see Introduction) scaling of the mean square displacement
of a tagged particle in a system of hardcore interacting particles and
how the dynamics of reaction-diffusion systems are modified as the
potential height $\phi_0$ is made finite.

%%%%%%%%%%%%%%%%%%%%%%%%%%%%%%%%%%%%%%%%%%%%%%%%%%%%%%%%%%%%%%%%%%%%%
% A C K N O W L E D G E M E N T S

\section{Acknowledgments}

T.A. acknowledges the support from the Knut and Alice Wallenberg
Foundation. Part of this research was supported by the NSF under grant
CHE0556268.

\appendix

\section{Stochastic simulations}\label{sec:Gillespie}

Stochastic simulations using the Gillespie algorithm
\cite{DG1,DG2,DG3} is a technique well suited for generating
stochastic trajectories for interacting particles systems. Briefly,
(similarly to
Ref. \onlinecite{Lizana_Ambjornsson,Novotny,Ambjornsson_PRL,Banik}) we
consider hopping of two particles on a one-dimensional lattice, with
lattice constant $a$. The number of lattice sites is
denoted by $M$ and chosen sufficiently large so that the ends of the
lattice are not reached. The dynamics is governed by the 'reaction'
probability density function
\be\label{eq:RPDF}
P(\tau,\mu)=k_\mu \exp(- \sum_\mu k_\mu
\tau) 
\ee 
where $\tau$ is the waiting time between jumps and $k_\mu$ are the
corresponding jump rates. There are four jump rates for the two
particle system considered in this study: the rate for particle 1
jumping to the left (right) is $k_1$ ($k_2$); similarly we denote by
$k_3$ ($k_4$) the left (right) jump rates for particle 2. For the case
that the two particles are at a distance smaller or larger than half
the barrier width $b/2$, the two particles diffuse independently and
we set $k_1=k_2=k_3=k_4=k_{\rm free}$. If two particles are separated
by a distance $b/2$ we set reduced jump rates according to
$k_2=k_3=k_{\rm free}\exp(-\beta \phi_0)$. We generate a stochastic
time series through the steps: (1) place the particles at their
initial positions; (2) From the PDF given in Eq. (\ref{eq:RPDF})
we generate the random numbers $\tau$ (waiting time) and $\mu$ (what
particles to move and in what direction) using the direct method
\cite{DG1}; (3) Update the positions $X_i$ ($i=1,2$) of the particles,
the time $t$ and the rates $k_1,k_2,k_3$ and $k_4$ for the new
configuration and return to (1). This procedure produces a stochastic
time series $X_i(t)$. Steps (1)-(3) are repeated $n$ times ($n$
ensembles) in order to obtain the ensemble averaged overtake and
trapping probabilities or the tagged particles PDFs (which are the
entities calculated in the main text). The ensemble averaged results
of a Gillespie time series is equivalent to the solution of a master
equation incorporating the rates given above \cite{DG1,DG2}. In the
limit $a\rightarrow 0$ with fixed diffusion constant, $D=k_{\rm
free}a^2$, the master equation approaches the diffusion equation as
specified in section \ref{sec:ProblemDef}.

%%%%%%%%%%%%%%% Bibliography %%%%%%%%%%%%%%%%%%%% 

%\newpage

\end{document}